\documentclass[conference]{IEEEtran}
\IEEEoverridecommandlockouts
\usepackage{cite}
\usepackage{caption}
\usepackage{subcaption}
\usepackage{amsmath,amssymb,amsfonts}
\usepackage{algorithmic}
\usepackage{graphicx}
\usepackage{tabularx}
\usepackage{textcomp}
\usepackage{xcolor}
\def\BibTeX{{\rm B\kern-.05em{\sc i\kern-.025em b}\kern-.08em
    T\kern-.1667em\lower.7ex\hbox{E}\kern-.125emX}}
\begin{document}

\title{Genetic Sequence compression using Machine Learning and Arithmetic Encoding Decoding Techniques\\
}

\author{\IEEEauthorblockN{1\textsuperscript{st} Mehedi Hasan Sarkar}
\IEEEauthorblockA{\textit{Department of CSE} \\
\textit{Stamford University Bangladesh}\\
Dhaka, Bangladesh \\
mehedihasansarkar1899@gmail.com}
\and
\IEEEauthorblockN{2\textsuperscript{nd} Adnan Ferdous Ashrafi}
\IEEEauthorblockA{\textit{Department of CSE} \\
\textit{Stamford University Bangladesh}\\
Dhaka, Bangladesh \\
adnan@stamforduniversity.edu.bd}
}

\maketitle

\begin{abstract}
We live in a period where bio-informatics is rapidly expanding, a significant quantity of genomic data has been produced as a result of the advancement of high-throughput genome sequencing technology, raising concerns about the costs associated with data storage and transmission. The question of how to properly compress data from genomic sequences is still open. Previously many researcher proposed many compression method on this topic DNA Compression without machine learning and with machine learning approach. Extending a previous research, we propose a new architecture like modified DeepDNA and we have propose a new methodology be deploying a double base-ed strategy for compression of DNA sequences. And validated the results by experimenting on three sizes of datasets are 100, 243, 356. The experimental outcomes highlight our improved approach's superiority over existing approaches for analyzing the human mitochondrial genome data, such as DeepDNA.
\end{abstract}


\section{Introduction}
We live in a period where bio-informatics is rapidly expanding, for example, to support precision medicine's unique treatment of each patient based on sequencing. However, sequencing data for a single individual might need terabytes, requiring enormous databases. By optimizing data compression techniques, this can result in significant cost reductions\cite{b1}. The rate at which DNA is being sequenced is exponentially growing, pushing the limits of genomics storage. There will be a yearly storage need of between 2 and 40 EB by the year 2025, according to some forecasts\cite{b2}.\\
Lossy or lossless compression techniques are available at the moment of compression. Lossless data compression algorithms are typically employed to carry out archive-related tasks or any other high-quality tasks.
There are many lossless algorithm available. The Huffman Coding, Run Length Encoding, Arithmetic Encoding, and Dictionary Based Encoding are some of the principal methods used\cite{b3}.\\
Arithmetic coding outperforms the more well-known Huffman approach in most ways. It expresses information at least as compactly, if not more so. Its performance is ideal without the requirement for input data blocking. It promotes a clear distinction between the model for expressing data and the information encoding in relation to that model. It can readily accept adaptive models and is computationally efficient. Nonetheless, many writers and practitioners appear to be ignorant of the approach. Indeed, many people believe that Huffman coding cannot be improved\cite{b4}.\\
In this paper Arithmetic DNA encoding using as compression algorithm. For DNA base probability table is using arithmetic encoding, it created using Deep Learning Technique.\\
Deep learning is a machine learning approach that trains computers to do what people do instinctively: learn by doing. Deep learning is a major technology underpinning self-driving cars, allowing them to detect a stop sign or discriminate between a pedestrian and a lamppost. It is essential for voice control in consumer electronics such as phones, tablets, televisions, and hands-free speakers. Deep learning has received a lot of attention recently, and for good reason. It is attaining outcomes that were previously unthinkable.\\
A computer model learns to execute categorization tasks directly from pictures, text, or sound in deep learning. Deep learning models may attain cutting-edge accuracy, sometimes outperforming humans. Models are trained utilizing a huge quantity of labeled data and multi-layered neural network architectures\cite{b6}.

\section{Literature Review}
In order to properly analyze the project, we thoroughly investigated any relevant work that had been done in the prior 12 years. We gathered the information and methods they suggested and disclosed. We came up with a new strategy and notion after gathering general principles that will result in a better data storage system that is trouble-free and takes less time.
\subsubsection{Human mitochondrial genome compression using machine learning techniques}
In this study, the authors suggest a method for compressing data from the human mitochondrial genome using machine learning techniques (DeepDNA). The compression effect of their method is comparable to that of reference-based methods, despite the fact that it is not. Additionally, their method produces well-compressed results in single genomes with little redundancy and population genomes with high redundancy.\cite{b5}.
\subsubsection{Context binning, model clustering and adaptivity for data compression of genetic data}
The context binning and model clustering are proposed as automated optimizations of Markov-like models in this work. They enable the optimization of a small number of models (as cluster centroids) to be selected, for example, separately for each read. A few adaptivity strategies that take into account data non-stationarity are briefly described.\cite{b1}.
\subsubsection{Efficient DNA sequence compression with neural networks}
In five datasets, including a balanced and comprehensive dataset of DNA sequences, the Y chromosome and human mitogenome, two collections of archaeal and viral genomes, four whole genomes, and two collections of FASTQ data of a human virome and ancient DNA, authors evaluate GeCo3 as a reference-free DNA compressor. They benchmark GeCo3 in 4 datasets made up of the pairwise compression of the chromosomes of the genomes of several primates in order to assess its effectiveness as a reference-based DNA compressor. GeCo3 is a genomic sequence compressor that outperforms other genomic compressors by utilizing a neural network mixing technique.\cite{b2}.
\subsubsection{A New Approach towards Compression of a Genetic Sequence using Unique Pattern Indexing and Mining Frequent Pattern}
Author propose a fast and efficient algorithm for mining frequent patterns in multiple DNA sequences. Two algorithms for this paper based on Unique Pattern Indexing and Searching Frequent Pattern. The first one is DNA sub-sequence replaced by unique index. The second algorithm is Sorting and replacing Sorting ID with Index ID. The experimental results shown that their proposed approach is memory efficient and mines maximal contiguous frequent patterns within a minimum time and reduce typical memory usage by 57.64\% at the very minimum. Author got a tremendous success in the field of bio-informatics. This is really useful and will give a remarkable changes in big data storage system.\cite{b7}.
\subsubsection{Rewritable two-dimensional DNA-based data storage with machine learning reconstruction}
This article describes a two-dimensional molecular data storage system that performs nontrivial joint data encoding, decoding, and processing while storing information in both the sequence and the backbone structure of DNA. They create machine learning algorithms for automatic discolouration detection and picture in painting in order to prevent expensive worst-case redundancy for repairing sequencing/rewriting mistakes and to alleviate problems related to mismatched decoding parameters. By recreating a library of photos with minimal or no visual deterioration following readout processing, as well as by removing and rewriting copyright metadata embedded in nicks, the 2DDNA platform is experimentally evaluated. Their findings show that heterogeneous data may be stored in DNA as a write-once and rewritable memory and that data can be permanently wiped while maintaining privacy.\cite{b8}.
\subsubsection{A Reference-Free Lossless Compression Algorithm for DNA Sequences Using a Competitive Prediction of Two Classes of Weighted Models}
A novel lossless compressor with enhanced compression capabilities for DNA sequences representing various domains and kingdoms is described by the author in this study. Prior to applying arithmetic encoding, the reference-free technique estimates, for each symbol, the optimal class of models to be utilized using a competitive prediction model. Specific sub-programs are used by both kinds of models to effectively handle inverted repetitions. The findings demonstrate that, utilizing a competitive level of computational resources and a balanced and diversified benchmark, the suggested strategy achieves a greater compression ratio than state-of-the-art approaches\cite{b9}.
\subsubsection{Compression of deep neural networks: bridging the gap between conventional-based pruning and evolutionary approach}
Convolutional neural network model compression is presented in this study as a multiobjective optimization problem with the competing goals of performance improvement and model size reduction. To solve this issue, a brand-new structured pruning technique called Conventional-based and Evolutionary Approaches Guided Multiobjective Pruning (CEA-MOP) is developed, in which the effectiveness of conventional pruning techniques for the evolutionary process is fully used. A multiobjective optimization evolutionary model has automated the difficult balancing act between pruning rate and model correctness. In order to create a codebook for further evolutionary processes, an ensemble framework first combines pruning measures. The length of the chromosome is then reduced using an effective coding technique, guaranteeing its better scalability.Last but not least, sensitivity analysis is automatically performed to establish the maximum pruning rate for each layer. Notably, CEA-MOP increases relative accuracy and lowers FLOPs on ResNet-110 by more than 50\% on CIFAR-10.\cite{b10}.

\section{Methodology}
The architecture of our neural network model modified DeepDNA for single base and double base are shown in Fig-\ref{single-layer} and Fig-\ref{double-layer} respectively. The framework primarily consists of six layers, the first of which is a single one-hot representation that turns the nucleotides in the genomic sequence A, C, G, and T (single base) and AA, AC, AG, AT, CA, CC, CG, CT, GA, GC, GG, GT, TA, TC, TG, TT, A, C, G and T (double base) are sequentially represented in a, b, c, d, e, f, g, h, i, j, k, l, m, n, o, p, q, r, s and t into vectors. The context short-term correlation in the genome is extracted by the second convolution layer. The pooling layer, which is the third layer, helps to reduce noise. The fourth layer, LSTM (uses three LSTM layer in model), extracted the long-term correlation in the genome. Probabilities for the following nucleotides A, C, G, and T  are output for single base and a, b, c, d, e, f, g, h, i, j, k, l, m, n, o, p, q, r, s and t for double base using the completely linked layer and the final layer.
\subsection{Convolutional Neural Network (CNN)}
\begin{figure*}[htbp]
\centerline{\includegraphics[width=0.9\textwidth]{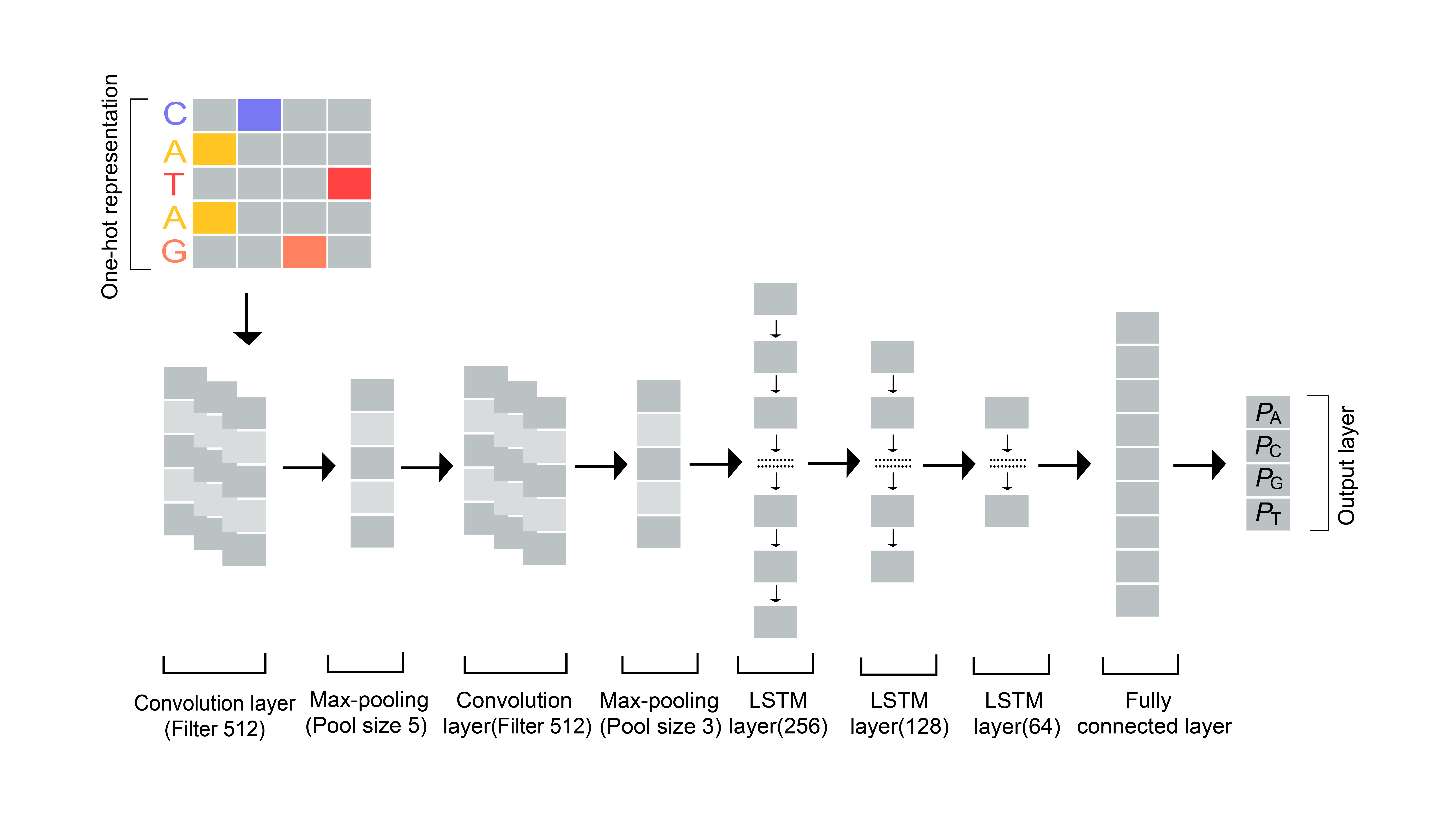}}
\caption{he architecture of the modified DeepDNA model for single base}
\label{single-layer}
\end{figure*}

\begin{figure*}[htbp]
\centerline{\includegraphics[width=0.9\textwidth]{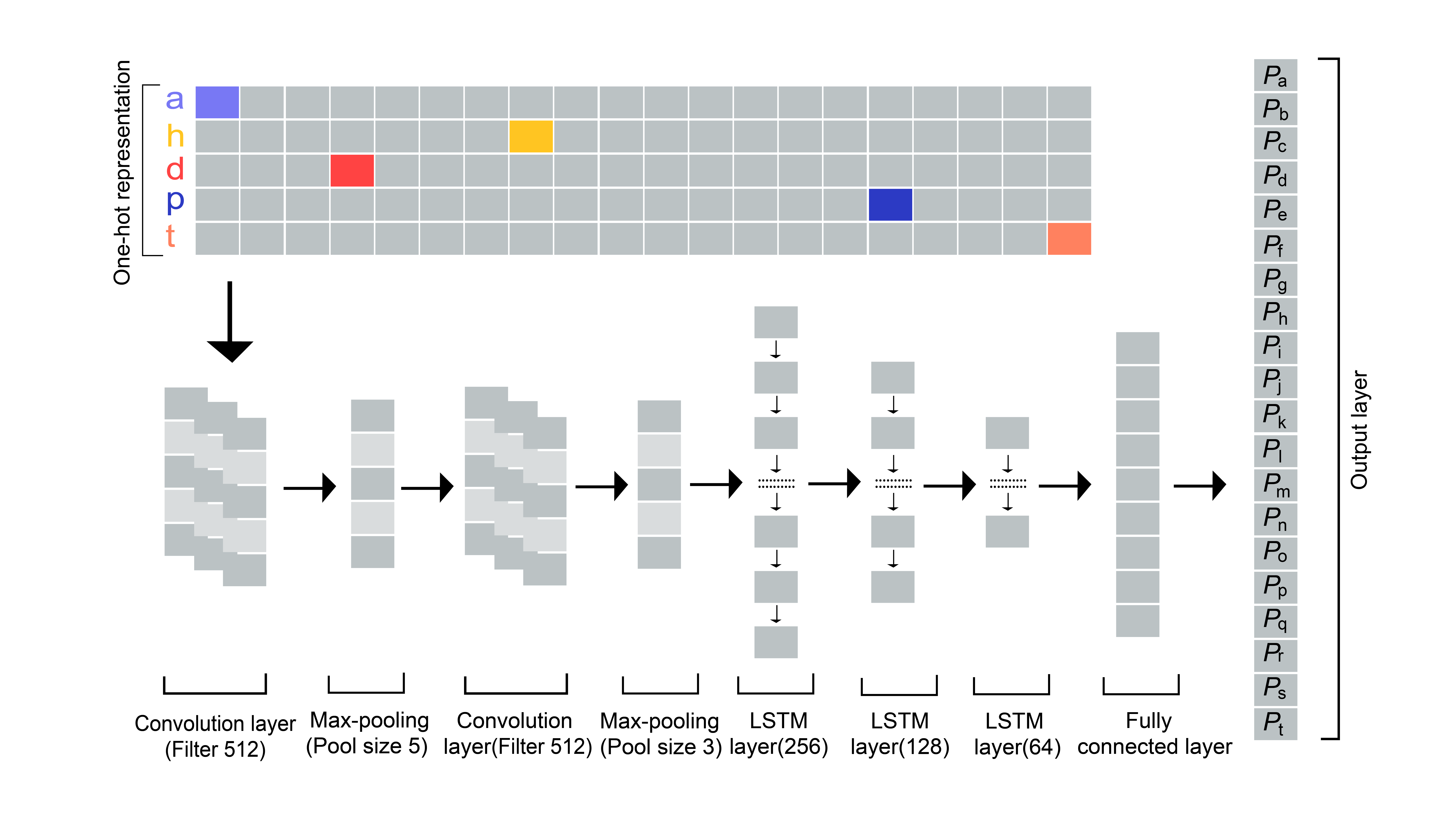}}
\caption{The architecture of the modified DeepDNA model for double base.}
\label{double-layer}
\end{figure*}

One-dimensional convolution operation uses the one-hot encoding of the genome as input and identifies the sequence characteristics at different points in the genome by sliding a series of filters across the sequence. One can be described in Fig-\ref{single-layer} and Fig-\ref{double-layer}.

A convolution layer activated by a rectified linear units acts as a local feature extractor, its output is a matrix with column matrix of the convolution filter and the row matrix of the position in the input sequence; A max-pooling procedure is used to reduce the size of the output matrix and only preserve the main features (Convolution layer and Max-pooling layer are repeated two times); The subsequent Long Short-Term Memory network (LSTM) layer is considered as acting the role of capturing sequence long-term features (Three LSTM layer uses here they are 256, 128, 64); A flattened fully connected layer is to collect LSTM outputs; The last layer performs a softmax non-linear transformation to a vector that serves as probability predictions of the sequence base
For single base A, T, C, G:\\
\begin{equation}
    \begin{split}
        A \to [1,0,0,0]\\
        C \to [0,1,0,0]\\
        G \to [0,0,1,0]\\
        T \to [0,0,0,1]\\
    \end{split}
\end{equation}  
For double base a, b, c, d, e, f, g, h, i, j, k, l, m, n, o, p, q, r, s and t:\\
\begin{equation}
    \begin{split}
        a \to [1,0,0,0,0,0,0,0,0,0,0,0,0,0,0,0,0,0,0,0]\\
        b \to [0,1,0,0,0,0,0,0,0,0,0,0,0,0,0,0,0,0,0,0]\\
        c \to [0,0,1,0,0,0,0,0,0,0,0,0,0,0,0,0,0,0,0,0]\\
        d \to [0,0,0,1,0,0,0,0,0,0,0,0,0,0,0,0,0,0,0,0]\\
        e \to [0,0,0,0,1,0,0,0,0,0,0,0,0,0,0,0,0,0,0,0]\\
        f \to [0,0,0,0,0,1,0,0,0,0,0,0,0,0,0,0,0,0,0,0]\\
        g \to [0,0,0,0,0,0,1,0,0,0,0,0,0,0,0,0,0,0,0,0]\\
        h \to [0,0,0,0,0,0,0,1,0,0,0,0,0,0,0,0,0,0,0,0]\\
        i \to [0,0,0,0,0,0,0,0,1,0,0,0,0,0,0,0,0,0,0,0]\\
        j \to [0,0,0,0,0,0,0,0,0,1,0,0,0,0,0,0,0,0,0,0]\\
        k \to [0,0,0,0,0,0,0,0,0,0,1,0,0,0,0,0,0,0,0,0]\\
        l \to [0,0,0,0,0,0,0,0,0,0,0,1,0,0,0,0,0,0,0,0]\\
        m \to [0,0,0,0,0,0,0,0,0,0,0,0,1,0,0,0,0,0,0,0]\\
        n \to [0,0,0,0,0,0,0,0,0,0,0,0,0,1,0,0,0,0,0,0]\\
        o \to [0,0,0,0,0,0,0,0,0,0,0,0,0,0,1,0,0,0,0,0]\\
        p \to [0,0,0,0,0,0,0,0,0,0,0,0,0,0,0,1,0,0,0,0]\\
        q \to [0,0,0,0,0,0,0,0,0,0,0,0,0,0,0,0,1,0,0,0]\\
        r \to [0,0,0,0,0,0,0,0,0,0,0,0,0,0,0,0,0,1,0,0]\\
        s \to [0,0,0,0,0,0,0,0,0,0,0,0,0,0,0,0,0,0,1,0]\\
        t \to [0,0,0,0,0,0,0,0,0,0,0,0,0,0,0,0,0,0,0,1]\\
    \end{split}
\end{equation}

\subsection{Arithmetic encoder}
The arithmetic encoder\cite{b11}, which Rissanen and Pasco introduced in 1976 to address the issue of infinite decimal accuracy, is a coding strategy that is most similar to information entropy under fixed data distribution. The arithmetic encoder encrypts the sequence as a sufficiently accurate numeric value in the range of (0, 1), known as the sequence identifier or label, as opposed to encoding each letter to an integer. The label interval gets smaller and smaller as the encoding goes forward, and the probability of an encoding character determines the next interval range.\\
The decoding process is identical to encoding in that the arithmetic decoder predicts the probability to the relevant character interval given the character probabilities. The decoder can recover the whole encoding sequence losslessly as long as the decimal representation of the interval identifier or label is precise enough.\\
\begin{figure}[htbp]
\centerline{\includegraphics[width=0.5\textwidth]{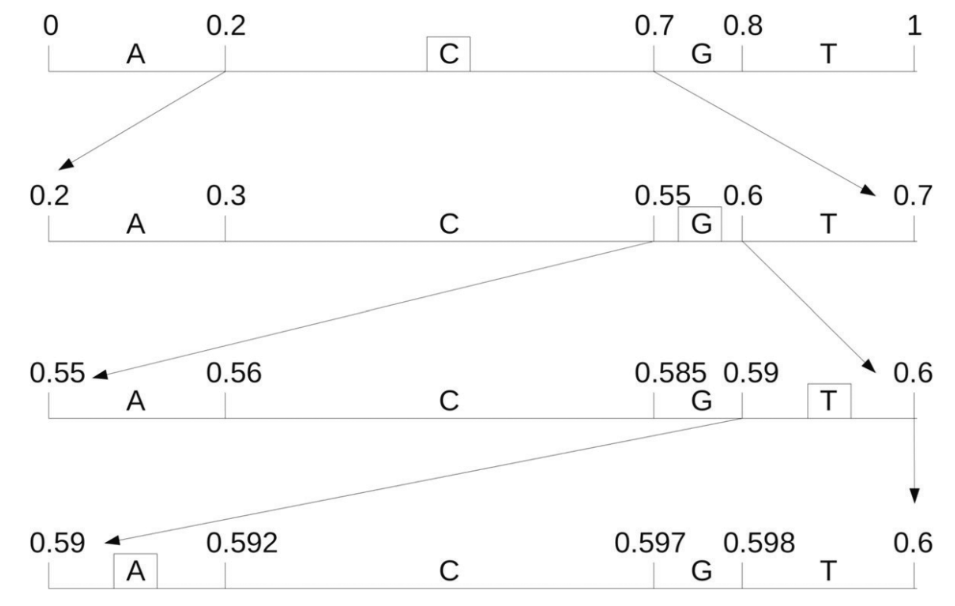}}
\caption{Arithmetic encoding process for single base DNA sequence}
\label{single-arithmetic}
\end{figure}
Fig-\ref{single-arithmetic} demonstrates the identifier determination procedure of arithmetic encoder, illustrates the sequence label determination process when encoding a sequence ‘CGTA’, we assume that the probability values of each base are: p(A) = p(T) = 0.2, p(C) = 0.5, p(G) = 0.1. At beginning, the initial interval is (0, 1), then the first base ‘C’ limited the interval to (0.2, 0.7), base ‘G’ limited the interval to (0.55, 0.6), and so on $\ldots$ . The latter interval is a subset of the former interval, so the range will be more  and more smaller, lastly, the identifier is limited the interval to [0.59, 0.592]. We can choose any point within this  interval, like a middle value 0.591, its binary value stream is the arithmetic coded description of the raw sequence ‘CGTA’.
The decoding process is similar to encoding. First, base ‘C’ is decoded according to 0.591 within the interval (0.2,0.7), while the next decoding process 0.591 within the interval (0.55,0.6), the base ‘G’ is decoded, and so on $\ldots$ , until decoded the last base ‘A’, and the whole sequence ‘CGTA’ is decoded.\\
\begin{figure}[htbp]
\centerline{\includegraphics[width=0.5\textwidth]{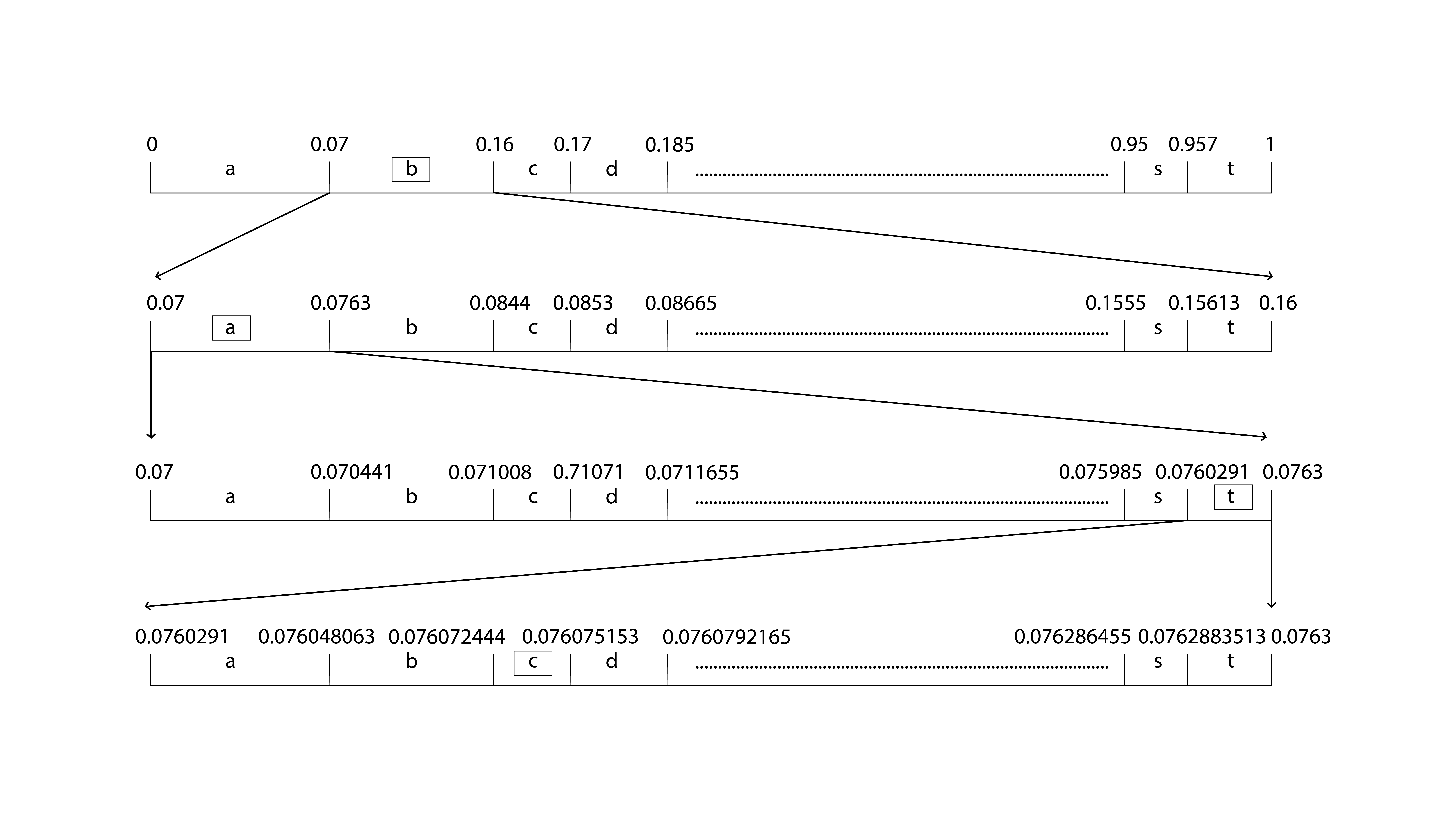}}
\caption{Arithmetic encoding process for double base DNA sequence}
\label{double-arithmetic}
\end{figure}
Fig-\ref{double-arithmetic} demonstrates the identifier determination procedure of arithmetic encoder, illustrates the sequence label determination process when encoding a sequence ‘batc’, we assume that the probability values of each base are: p(a) = 0.07, p(b) = 0.09, p(c) = 0.01, p(d) = 0.015 \ldots p(s) = 0.007, p(t) = 0.043. At beginning, the initial interval is (0, 1), then the first base ‘b’ limited the interval to (0.07, 0.16), base ‘a’ limited the interval to (0.07, 0.0763), and so on $\ldots$ . The latter interval is a subset of the former interval, so the range will be more  and more smaller, lastly, the identifier is limited the interval to [0.076072444, 0.076075153]. We can choose any point within this  interval, like a middle value 0.076073565, its binary value stream is the arithmetic coded description of the raw sequence ‘batc’.
The decoding process is similar to encoding. First, base ‘b’ is decoded according to 0.076073565 within the interval (0.07,0.16), while the next decoding process 0.076073565 within the interval (0.07,0.0763), the base ‘a’ is decoded, and so on $\ldots$ , until decoded the last base ‘c’, and the whole sequence ‘batc’ is decoded.\\
Consider an input sequence $x_{i-1}, x_{i-2}, \ldots , x_{i-k+1}$ for the compression model, the output for the model is the predict the next base xi. The estimate probabilities  $p(x_i|(x_{i-1}, x{i-2}, \ldots , x_{i-k+1}), h_{i-1})$ is provided into arithmetic encoder to obtain the final output bit-streams,  where $h_{i-1}$ is the deep learning model former state. According to the theory of Shannon entropy \cite{b12}, the number of output bits for the base $x_i$ is determined by:
\begin{equation}
    \begin{split}
        H = -\log_2(p(x_i|(x_{i-1},x_{i-2},\ldots,x_{i-k+1})),h_{i-1})
    \end{split}
\end{equation}

That is, the more accurate the probability of our model estimation, the higher the probability of corresponding coding base, the smaller the output bit-stream, and the better the compression effect we get.
\subsection{Model setting \& training}
We fully utilize the long-term feature extraction capability of LSTM and the local feature capture capability of CNN in the deep learning model. Our deep learning model is implemented using the Keras framework, an open deep learning resource developed from Theano's backend.
In the deep learning model, all parameters are initialized according to random and uniform distribution  unit (-0.05, 0.05), and entire biases are initialized to 0. We use batch training (default size: 4096) to minimize the  cross entropy loss function on the training data set. Validation losses were assessed at the end of each training  epoch to monitor the convergence. We utilized about 100 epochs to complete the training, each of which took about ~1 mins for single base and we utilized about 60 epochs to complete the training, ~40 sec for double base. They both are trained for 100 DNA sequence.\\
The loss function of cross-entropy is defined as:
\begin{equation}
    \begin{split}
        L(y,\widehat y) = - \frac{1}{n} \sum_{t=1}^n \sum_{i=1}^4 y_i^{(t)} \log(\widehat y_i^{(t)})
    \end{split}
\end{equation}
Where $\widehat y_i^{(t)}$ is the probability of prediction character at time $t$ being nucleotide $i, y_i^{(t)}$ is the one-hot vector repre- sent the real nucleotide at time $t$, and mini-batches sample  size is $n$. We utilized the adaptive various decay learning schedule in different experiment which used by RMSprop as the learning rate of the model.\\
For Both Single base and Double base DeepDNA, Both in the encoding and decoding process, it calculated  the nucleotide probability based on the same deep learning network parameters, so they get the same prediction  probability value, therefore, the original sequence can be lossless reconstructed by arithmetic coding. 

\section{Experiment and Result Analysis}

\subsection{Datasets}
One hundred entire human mitochondrial genome sequences were employed as experimental data to confirm the efficacy of our suggested model. The average length of the human mitochondrial genome sequence is 16,500 base pairs (bp).All data were download from the  MITOMAP \cite{b13} database in the 2021. For the training data set, we randomly picked 70 percent sequences, for the verification data set, 20 percent sequences, and for the test data set, 10 percent sequences. In order to make the sequence consists of only 4 nucleotides (A, C, G, T), we uses consensus approach for fuzzy symbols in the training validation process and all lowercase nucleotides in the data set were converted to uppercase.In the compression process, we record the nucleotides not in the {A,C, G, T} and its position information, record whether the base is in lowercase or not, so that the original sequence can be reconstructed in lossless when decompressing.
\subsubsection{Data ready for double base}
After the consensus, the sequence of DNA base change with double base into single base a, b, c, d, e, f, g, h, i, j, k, l, m, n, o, p, q, r, s, t and store it fasta file similar to previous mitochondrial genome sequence fasta file.

\subsection{Experimental Setup}
At first, data-sets are processed for both single base and double base. In single base modified DeepDNA has 4 base A, C, G, T. And in the double base modified DeepDNA has 20 base a, b, c, d, e, f, g, h, i, j, k, l, m, n, o, p, q, r, s, t. The sequence of DNA base A, C, G, T and a, b, c, d, e, f, g, h, i, j, k, l, m, n, o, p, q, r, s, t divided into three part. They are train dataset, validation dataset, test dataset.The training dataset was  used to learning model parameters, the verification dataset was used to determine network structure and model parameters, and the test dataset was used to verify the performance of the final selection of model parameters.\\
Then CNN, LSTM model are created for compress genetic sequence. Then the test data-sets are predicted and arithmetic encoding is applied with the prediction probability for test datasets. Encoding 64 base at a time and store it's decimal value in a file. When it is decoded the decimal value with the prediction probability for test datasets, then it get the exact same base which it encoded. This compression is 100 percent lossless compression.
\subsection{Result Analysis}

\begin{table}[htbp]
\centering
    \caption{Single base for different sequence learning rate Decay table}
    \label{tab:single-decay-table-100}
    
\small
    \begin{tabularx}{\columnwidth}{@{}XXXXXX}\hline
    Genome Sequences & Exponential Decay & Cosine Decay & Polynomial Decay & Inverse Time Decay & Cosine Restart Decay\\\hline\hline
    100 & 0.0083  & 0.0082  & 0.0084  & 0.0080 &\textbf{0.0078} \\
    243 & \textbf{0.0068}  & 0.0098  & 0.0092  & 0.0072 & - \\
    356 &  \textbf{0.0114} & 0.0154 & - & - & - \\\hline 
    \end{tabularx}\\
\end{table}

\begin{table}[htbp]
\centering
    \caption{Compression between DeepDNA Model and Modified DeepDNA Model (differnt learning Decay) for 100 DNA sequence (use same Dataset) for single base}
    \label{tab:compression-100}
    
\small
    
    \begin{tabularx}{\columnwidth}{@{} *{6}{c} @{}}\hline
    
    \multicolumn{1}{@{}c}{DeepDNA} &
\multicolumn{5}{c@{}}{Modified DeepDNA Model} \\
    \multicolumn{1}{@{}c}{Model} &
\multicolumn{5}{c@{}}{} \\
     &  Exponen- &  Cosine &  Poly- &  Inverse &  Cosine \\
     &  tial &  Decay &  nomial &  Time &   Restart\\
     &  Decay &   &  Decay &  Decay &   Decay\\\hline\hline
    0.0126 & 0.0083  & 0.0082  & 0.0084  & 0.0080 &\textbf{0.0078} \\\hline 
    \end{tabularx}\\
    \hbox{}
\end{table}

\begin{table}[htbp]
\centering
    \caption{Double base for 100 sequence different learning rate Decay table}
    \label{tab:double-decay-table-100}
    \begin{tabular}{p{3cm}p{3cm}}\hline
    Decay Name &  Bit Per Base (BPB) \\\hline\hline
    Exponential Decay & 0.0188 \\
    Cosine Decay & 0.0195 \\
    Polynomial Decay & 0.0188 \\
    Inverse Time Decay &\textbf{0.018798}  \\
    Cosine Restart Decay & 0.0192 \\\hline
    \end{tabular}\\
\end{table}
In Table-\ref{tab:single-decay-table-100}, five learning rate decay are experimented for 100, 243, 356 mitochondrial genome sequences for single base data. And Cosine restarts decay  result is best in 100 mitochondrial genome sequence. Table- \ref{tab:compression-100}, Modified DeepDNA model's all learning decay are good enough from DeepDNA Model with same data set for 100 genome sequences. And Exponential decay  result is best in 243 mitochondrial genome sequence. Other results are also better from previous working DeepDNA model. And Table-\ref{tab:double-decay-table-100}, five learning rate decay are experimented for 100 mitochondrial genome sequence in double base data. And Inverse Time decay  result is best. Other result are also better.\\

Modified DeepDNA corresponds to the average of the lengths of all base prediction probability output codes in the genome sequence, namely:
\begin{equation}
    modified DeepDNA(bpb) = - \frac{1}{T} \sum_{i=1}^T \log_2 (P(\widehat y_i))
\end{equation}
Where $P(y_i)$ is the predicted probability value of the modified DeepDNA model output corresponding to the base $y_i$ of the genome sequence at the i-th position, and $T$ is the sequence length of the compression genome. These values can be directly fed into arithmetic coding \cite{b4} to produce the compressed bit-streams file.

\section{Conclusion}
The convolutional neural network (CNN) and the long short-term memory network (LSTM) are used in a revolutionary machine learning technique we modified a model called DeepDNA to compress genomic sequences. An experiment using 100, 243, 356 entire human mitochondrial genome sequences has demonstrated that our technique is capable of learning local aspects of sequences through convolution layers and sophisticated representations of sequence dependency using long short-term memory networks (LSTM). On a challenge that involved compressing 100, 243, 356 sequences from the human mitochondrial genome, we evaluated the deep learning model's performance and came up with a passable result.\\
Our model demonstrated the viability of using CNN and LSTM network models to compress genomic sequences. This effort will help to clarify the relationship between genes and illness, better investigate the patterns and laws in genome sequences, and help decode the functional properties of sequences. Because a stronger sequence prediction model will result in a greater compression effect, it will be able to help with all of the aforementioned issues.\\
Additionally, the compression approach may gather more redundant information and provide higher enhanced compression efficiency with the study of genome sequence properties in future biological analysis. 





\end{document}